\documentclass[manuscript]{aastex62}
\usepackage{times}
\usepackage{graphicx}
\usepackage{epstopdf}
\usepackage{url}
\usepackage{verbatim}
\usepackage{lineno}

\def\ergscm2 {erg\,s$^{-1}$cm$^{-2}$}

\def\cm2 {cm$^{-2}$}

\shorttitle{\textsc{The event rates of low- and high-luminosity GRBs}}
\shortauthors{\textsc{Dong et al. 2023}}

\begin{document}

\title{THE ORIGIN OF LOW-REDSHIFT EVENT RATE EXCESS AS REVEALED BY THE LOW-LUMINOSITY GRBS}

\author{X. F. Dong}
\affiliation{College of Physics and Physical Engineering, Qufu Normal University, Qufu, 273165, P. R. China}
\author{Z. B. Zhang}\thanks{Email:z-b-zhang@163.com}
\affiliation{College of Physics and Physical Engineering, Qufu Normal University, Qufu, 273165, P. R. China}
\affiliation{Department of Physics, College of Physics, Guizhou University, Guiyang, 550025, P. R. China}
\author{Q. M. Li}
\affiliation{Department of Physics, College of Physics, Guizhou University, Guiyang, 550025, P. R. China}
\author{Y. F. Huang}\thanks{Email:hyf@nju.edu.cn}
\affiliation{School of Astronomy and Space Science, Nanjing University, Nanjing 210023, P. R. China}
\affiliation{Key Laboratory of Modern Astronomy and Astrophysics (Nanjing University),\\ Ministry of Education, Nanjing 210023, P. R. China}
\author{K. Bian}
\affiliation{College of Physics and Physical Engineering, Qufu Normal University, Qufu, 273165, P. R. China}

\begin{abstract}
The relation between the event rate of long Gamma-Ray Bursts and the star
formation rate is still controversial, especially at the low-redshift end.
Dong et al. confirmed that the Gamma-Ray Burst rate always exceeds the
star formation rate at low-redshift of $z<1$ in despite of the sample
completeness. However, the reason of low-redshift excess is still unclear.
Since low-luminosity bursts are at smaller redshifts generally, we choose
three Swift long burst samples and classify them into low- and
high-luminosity bursts in order to check whether the low-redshift excess
is existent and if the excess is biased by the sample size and completeness.
To degenerate the redshift evolution from luminosity, we adopt the
non-parametric method to study the event rate of the two types of long
bursts in each sample. It is found that the high-luminosity burst rate is
consistent with the star formation rate within the whole redshift range
while the event rate of low-luminosity bursts exceeds the star formation
rate at low redshift of $z<1$. Consequently, we conclude that the
low-redshift excess is contributed by the low-luminosity bursts with
possibly new origins unconnected with the star formation, which is also
independent of the sample size and the sample completeness.

\end{abstract}


\section{Introduction}
\label{sec:Intro}
Gamma-Ray Bursts (GRBs) can be traditionally classified into long and short
classes according to the duration ($T_{90}$), with a dividing line of $\sim 2$ s \citep[e.g.][]{1993ApJ...413L.101K,2008A&A...484..293Z,2020ApJ...893...77W} or $\sim1$ s \citep{2013ApJ...764..179B,2014ApJS..211...12G,2020ApJ...902...40Z}.
The exact dividing line between long and short GRBs depends on the sample and
the energy band. For example, the $T_{90}$ values of Swift GRBs with good
duration measurement \citep{2014ApJS..211...12G} and well-measured
spectrum \citep{2020ApJ...902...40Z,Deng2022} are bimodally distributed
with a boundary of $T_{90}\sim1$ s. In any case, it proves that the apparent
middle class in some $T_{90}$-selected GRB samples is inexistent or artificial \cite[e.g.][]{2016MNRAS.462.3243Z,2019ApJ...870..105T,2019ApJ...887...97T,2020ApJ...902...40Z,Deng2022,2022A&A...657A..13T}.
Long GRBs (lGRBs) with $T_{90}>2$ s are believed to come from the core-collapse of massive stars  \citep[e.g.][]{1993AAS...182.5505W,1998ApJ...494L..45P,2006ARA&A..44..507W,2003Natur.423..847H,2003ApJ...591L..17S,2018pgrb.book.....Z},
so that the lGRB rate was frequently compared with the star formation rate (SFR)
in a low-metallicity environment   \citep{2009MNRAS.400L..10W,2021A&A...649A.166P,2018RMxAA..54..309E}.
On the other hand, short GRBs (sGRBs) with $T_{90}<2$ s are usually interpreted as
the mergers of compact objects such as double neutron stars (DNS) or neutron
star-black hole (NS-BH) binaries \citep[e.g.,][]{1989Natur.340..126E,Paczynski1991,1992ApJ...395L..83N}.
lGRBs can also be divided into high- and low-luminosity classes \cite[see][for a detail]{2018pgrb.book.....Z}.
They can even be grouped into classical GRBs, X-ray-rich GRBs and
X-ray flashes (XRFs) \citep{2020ApJ...902...40Z} by considering their
peak energy, or into more interesting sub-classes by considering whether
they have extended emissions or not \citep{2020RAA....20..201Z}. Considering
the joint observational features, \cite{2009ApJ...703.1696Z} proposed to
classify GRBs into two subgroups of long/soft type I and short/hard type II in physics.

Since most lGRBs are believed to have a massive star core-collapse origin,
their redshift distribution is expected to trace the SFR, which has been
revealed by various SFR
indicators \citep[namely][]{1998ApJ...498..106M,2006MNRAS.372.1034D,2006ApJ...651..142H,2012MNRAS.423.3049G}.
Over the past decades, the event rates derived with different methods are still
controversial and even conflicting with the star formation history. Interestingly,
it was found that the lGRB rate exhibits an excess at
either the low-redshift end \citep[e.g.][]{2015ApJS..218...13Y,2015ApJ...806...44P,2018ApJ...852....1Z,2019MNRAS.488.5823L}
or the high-reshift end \cite[e.g.][]{2008ApJ...683L...5Y,2008ApJ...673L.119K,2011MNRAS.417.3025V,2019MNRAS.488.4607L}
when it is compared with the SFR. Since the current high-z SFR is measured
with a relatively large uncertainty \cite[see e.g.][hereafter D22]{2022MNRAS.513.1078D},
the discrepancy between the lGRB formation rate and the SFR at the high redshift end is not
prominent yet. On the contrary, whether the lGRB rate traces the SFR in the low-z domain
is still an important issue that has been deeply investigated recently \cite[e.g.][]{2015ApJS..218...13Y,2015ApJ...806...44P,2016A&A...587A..40P,2021A&A...649A.166P,2021ApJ...914L..40D,2022MNRAS.513.1078D,2022arXiv220601764N}. Similarly, some authors claimed that the formation rate of sGRBs also
exceeds the SFR at low redshift \citep{2018ApJ...852....1Z,2021ApJ...914L..40D}.
Recently, \cite{2022arXiv220601764N} analyzed the stellar population
of 69 sGRB host galaxies and concluded that the event rate of sGRBs trace
a combination of recent star formation and stellar mass, but is not singularly
dependent on either property. It is interesting to note that some authors argued that
low-luminosity GRBs are a distinctive class of
bursts \cite[e.g.][]{2004Natur.430..646S,2006ApJ...645L.113C,2007ApJ...662.1111L}.
However the relation between their event rate and the low-redshift lGRB rate
has not been carefully examined yet. Since most low-luminosity GRBs occur
at a relatively low redshift, it is quite possible that they could be the
key factor leading to the low-redshift excess of lGRB rate. Here
we will study this issue in-depth.

People have employed various statistical methods to assess the GRB formation
rate and compare it with SFR. There are mainly two approaches frequently applied
for this purpose. The first one is the traditional parametric estimate, namely the
direct fit \citep{2004ApJ...611.1005G,2005A&A...435..421G,2006MNRAS.372.1034D,2007ApJ...661..394L,2007ApJ...662.1111L,2008ApJ...683L...5Y,2010MNRAS.406.1944W,2015MNRAS.448.3026W,2018RMxAA..54..309E} or the maximum likelihood method \citep{2019MNRAS.488.4607L} on the stellar population analysis \citep{2022arXiv220601764N,2022arXiv220606390G},
in which the dependence of luminosity on redshift was neglected and the luminosity
distribution should be assumed in advance. The second one is the non-parametric
$\tau$-statistic method \citep{1971MNRAS.155...95L,1992ApJ...399..345E,2015ApJ...806...44P}
that was initially proposed to investigate the luminosity function of quasars and its evolution
based on a flux-limited sample. Noticeably, the evolution of luminosity with redshift can be
represented by a factor of $(1+z)^\delta$. It has now been popularly adopted to calculate the
luminosity function and event rate of both lGRBs and sGRBs \citep{2012MNRAS.423.2627W,2015ApJ...806...44P,2015ApJS..218...13Y,2016A&A...587A..40P,2017ApJ...850..161T,2018ApJ...852....1Z,2019MNRAS.488.5823L,2021ApJ...914L..40D,2022MNRAS.513.1078D}. It should be emphasized that the derived event rate could be somewhat biased by the
statistical method, the instrumental effect and the sample selection. By contrast, the
non-parametric method has a special advantage for a truncated sample since it does not
depend on any pre-assumptions of the luminosity property and can give point-by-point
description of the cumulative distributions. Therefore, we will use the non-parametric
$c^{-}$ method to investigate different samples of Swift lGRBs in this study.

To reduce the instrumental bias, GRBs detected by a single satellite are usually
utilized as the first choice. As a result, most previous works have paid particular
attention to the Swift/BAT GRBs. However, note that such GRB samples collected from
the literature are generally incomplete according to the selection standards defined
in \cite{2006A&A...447..897J} and \cite{2012ApJ...749...68S}. The completeness of GRB
sample was especially emphasized in \cite{2016A&A...587A..40P} (hereafter P16), but
our recent work showed that it does not affect the resulting event rate too much as
long as the selected GRB sample have a large size and a sufficient redshift
coverage \citep{2022MNRAS.513.1078D}. It strongly indicates that the low-redshift
excess of GRB rate is not due to the sample incompleteness. Also, even for the same
Swift lGRB sample, different methods may lead to inconsistent results. For example,
P16 used the non-parametric method to study a complete sample of 81 Swift lGRBs and
found no low-redshift excess, which however may be caused by the neglect of the
differential comoving volume in their calculations. Using the same complete lGRB sample
in P16, \cite{2019MNRAS.488.4607L} employed the maximum likelihood method to examine
the luminosity function and event rate and concluded that the GRB event rate is
roughly consistent with the SFRs at $z<2$ but shows a discrepancy at high redshift
of $z>2$. However, D22 also applied the non-parametric method to the P16 samples and
confirmed the similar low-redshift ($z<1$) excess and the high-redshift consistency
in both complete and non-complete Swift lGRB samples. Meanwhile, they concluded that
the resulting event rate is basically threshold-independent provided that the number
of bursts above the flux limit is sufficient enough. However, whether the event rate of low-luminosity GRBs differs from that
of high-luminosity ones and how these two types of GRBs trace star formation
are still unknown yet. It motivates us to utilize a larger Swift GRB sample to
investigate such an important issue in this study. Also, we will compare the
event rates derived from complete samples with that from incomplete samples.

This letter is arranged as follows. Section \ref{sec:sample} describes our sample selection,
data reduction and method. The detailed results are presented in Section \ref{sec:results}.
We summarize our findings in Section \ref{sec:summary} and end with a discussion in
Section \ref{sec:discussion}.

\section{DATA and method}
\label{sec:sample}

To obtain the less-biased luminosity function and event rates of low-
and high-luminosity lGRBs and their dependence on the redshift, we
need to collect those lGRBs with both redshift measurement and well-constrained spectra so that the bolometric luminosity can be precisely calculated. Firstly, we select 365 lGRBs with redshift and peak photon flux in
the Swift/BAT catalog\footnote{https://swift.gsfc.nasa.gov/archive/grb\_table/}
from January 26, 2005 to May 31, 2022. Note that the big lGRB sample
(called BL sample thereinafter) does not surfer from any selection
effects except the Malmquist bias \citep{Dainotti2018}. The method used to
convert the peak photon flux into peak energy flux ($F_p$) can be found
in \cite{2018PASP..130e4202Z}. Owing to the narrow energy range of the
Swift/BAT detector, a single power-law spectrum
of $\Phi(E)\propto E^{-\alpha}$ \cite[see also][]{2019ApJS..245....1T}
has been applied to calculate the bolometric
luminosity $L=4{\pi}d_{L}^{2}(z)F_pK_c$, where $F_p$ is in units of
erg\ cm$^{-2}$ s$^{-1}$, $K_c$ is the K-correction factor
and $d_{L}(z)=cH_0^{-1}(1+z)\int_0^z[\Omega_{\Lambda}+\Omega_{m}(1+z)^3]^{-0.5}dz$
denotes the luminosity distance at a redshift of z. Here, a flat $\Lambda$CDM
universe with ${\Omega}_{m}=0.27$, ${\Omega}_{\Lambda}=0.73$
and $H_{0}=70~{\rm km~s^{-1}Mpc^{-1}}$ will be used \citep{2021MNRAS.503.3262Z}.
Secondly, we also take the sample of 127 Swift lGRBs with well-measured
spectra from \cite{2015ApJS..218...13Y} (hereafter Y15) and the complete
sample of 79 lGRBs in D22 in order to check whether the resulting event
rates of lGRBs are dependent on the spectral quality, the sample size and
the sample completeness.

\begin{figure*}
  \centering
  \includegraphics[width=1.0\linewidth, angle=0,scale=0.9]{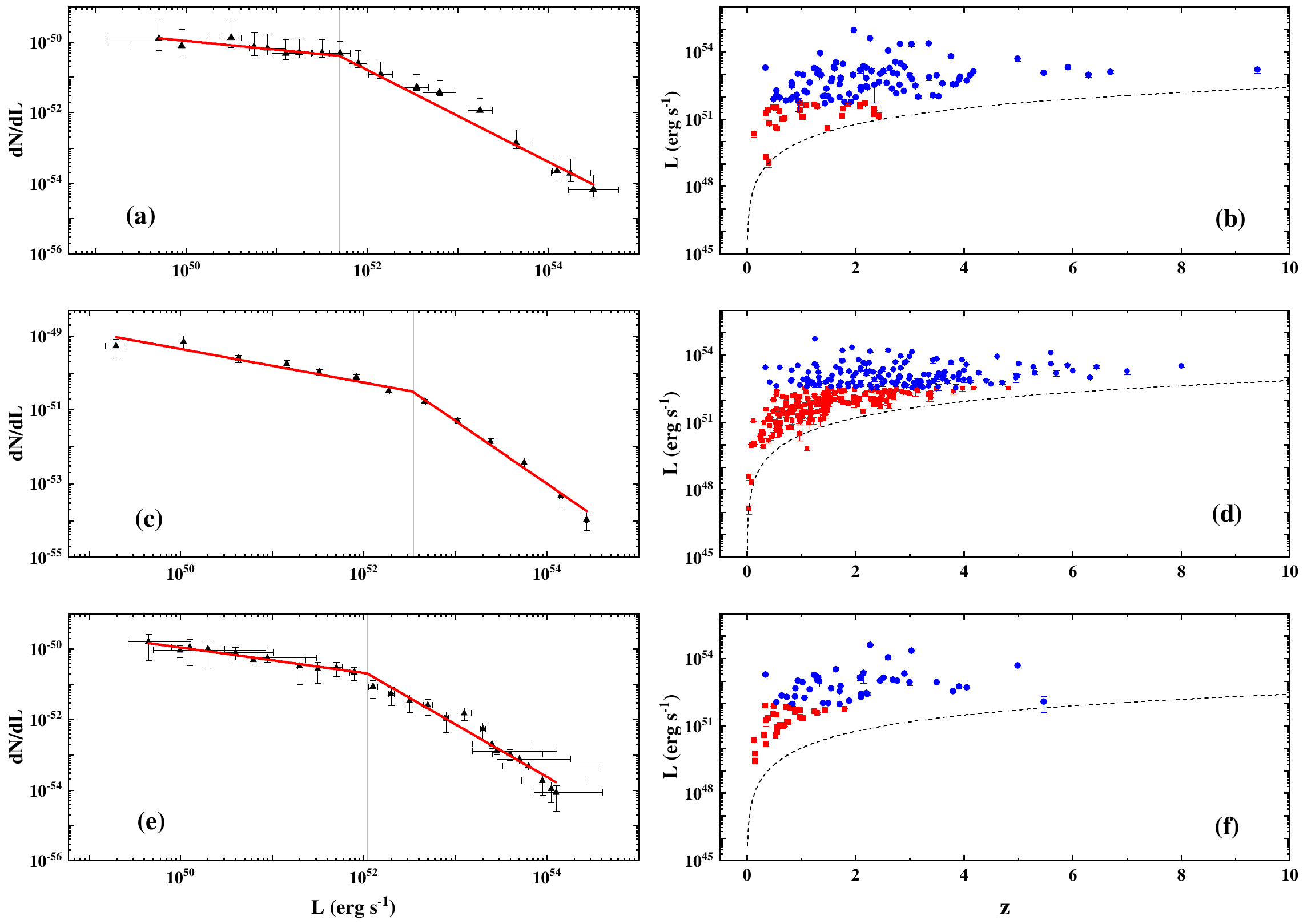}\\
  \caption{The differential luminosity distributions of Samples Y15, BL and D22,
  shown in Panels (a), (c) and (e), respectively. The solid thick lines denote
  the best fits with a broken power-law of Equation (\ref{broken power law})
  and the vertical thin lines mark the boundaries between the low- and
  high-luminosity events. Panels (b), (d) and (f) plot the luminosity versus
  reshift for the Y15, BL and D22 samples, correspondingly, together with
  the estimated threshold (dashed line). The observed low- and high-luminosity lGRBs
  are shown with filled squares and circles, respectively.  }
  \label{fig1}
\end{figure*}

\begin{figure*}
  \centering
  \includegraphics[width=1\linewidth, angle=0,scale=0.5]{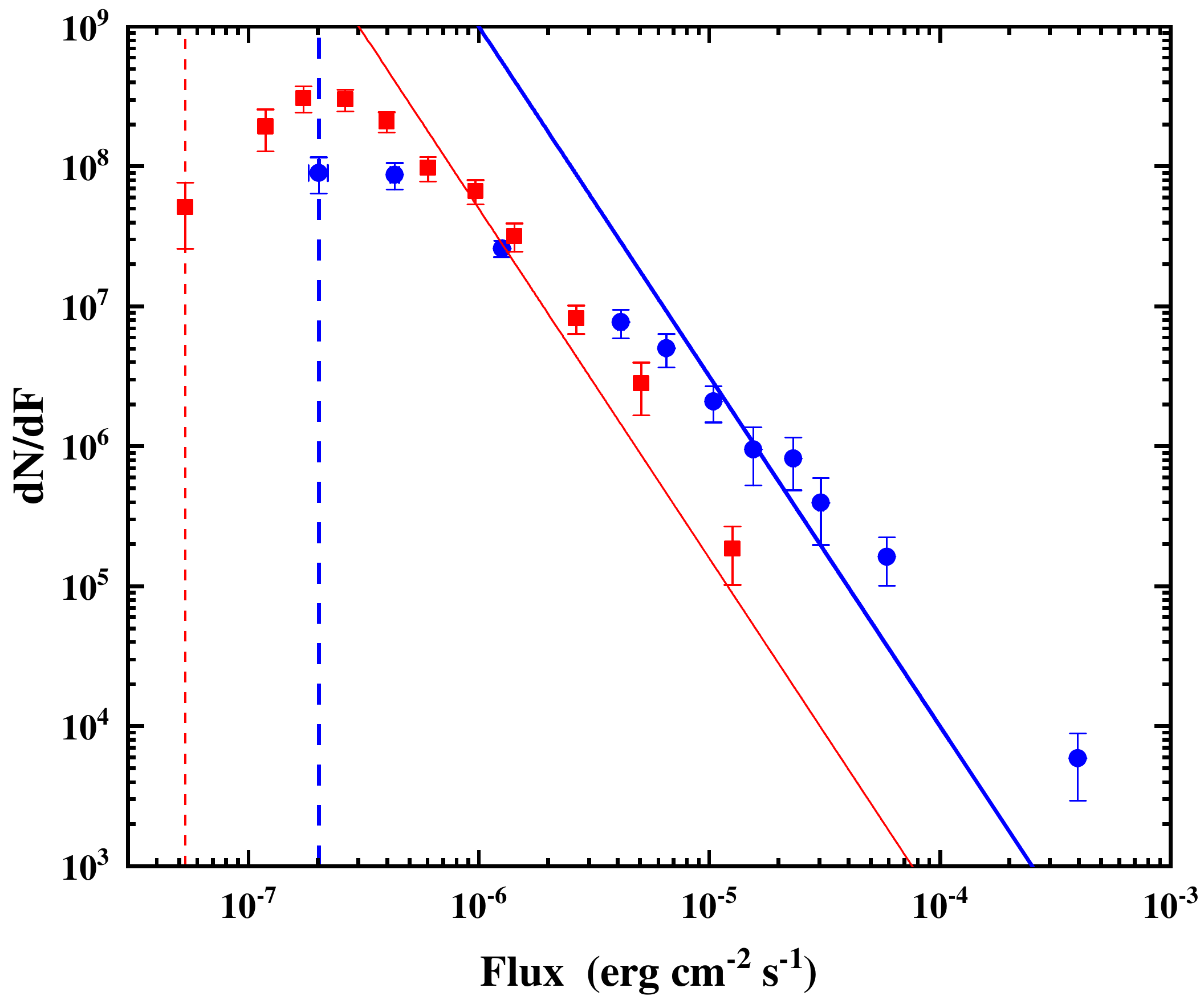}\\
  \caption{The differential flux distributions of low (filled squares)
  and high (filled circles) luminosity lGRBs. The two vertical dashed
  lines represent the smallest fluxes of them. The two solid lines
  denote the homogeneous distribution of $N\propto F^{-3/2}$ in the
  Euclidian space.}
  \label{fig2}
\end{figure*}

Subsequently, we divide the above three samples of lGRBs into low-luminosity
and high-luminosity ones according to the differential luminosity distributions
as shown in Figure \ref{fig1}. In this figure, the broken luminosity is marked
by a vertical line in the left panels, and the luminosity versus the redshift
of low- and high-luminosity lGRBs is displayed in the right panels.
To put the luminosity limit on the plots, we need to set a suitable flux
limit in advance. For the samples of Y15 and D22, we have taken the threshold
as $F_{lim}=2\times10^{-8}~{\rm erg~cm^{-2}s^{-1}}$. For our newly-built sample
BL, we plot the differential flux distributions of the two kinds of lGRBs in
Figure \ref{fig2}, from which we can get the smallest flux
as $F_{lim}=5.33\times10^{-8}~{\rm erg~cm^{-2}s^{-1}}$. This value is taken
as the threshold in order to reserve sufficient low-luminosity lGRBs.
It is worth noting that we have determined the boundaries between low-
and high-luminosities via the following broken power-law function
\begin{equation}\label{broken power law}
dN/dL\propto\left\{
\begin{array}{ll}
        (L/L_b)^{\alpha_1}, \ \
        & L\leq L_b \\
        (L/L_b)^{\alpha_2}, \ \
        & L>L_b
\end{array},
\right.
\end{equation}
which has been widely adopted in previous
studies \cite[e.g.][]{2002ApJ...574..554L, 2005A&A...435..421G, 2015ApJ...806...44P}.
Note that the broken power-law function presents a better description for the
luminosity function in Figure \ref{fig1}, as compared with the single power-law
function.

As a result, we get the broken luminosity as $L_b\sim4.92\times10^{51}$ erg s$^{-1}$,
$3.4\times10^{52}$ erg s$^{-1}$ and $1.09\times10^{52}$ erg s$^{-1}$ for the Y15,
BL and D22 samples, respectively, which are larger than the former artificial
reference value of $L_b\sim10^{50}$ erg s$^{-1}$ by about 1 -- 2 orders of magnitude.
In practice, there are no strict criteria to distinguish between low- and
high-luminosity lGRBs previously. For example, \cite{ 2006ApJ...645L.113C}
suggested that SN 1998bw/GRB 980425 and SN 2003lw/GRB 031203 are subluminous
in $\gamma$-rays but they did not provide a standard for the
classification \cite[see also][]{2007ApJ...662.1111L}. The fractions of
low-luminosity lGRBs on basis of our criterion given by
Equation (\ref{broken power law}) are $\sim$22\% for sample Y15,
$\sim$44\% for sample BL and $\sim$41\% for sample D22. For the sample
BL, we caution that the percentages of low-redshift $(0<z<1)$ lGRBs
are $\sim$ 36$\%$ and $\sim$6.2$\%$ for the low- and high-luminosity
lGRBs, respectively. The fraction of low-luminosity lGRBs goes up
to $\sim$88\% in the redshift range from 0 to 1. This demonstrates
that most low-luminosity lGRBs are located at high-redshift instead
of in the nearby universe and will be absolutely dominant at the
low-redshift end.

\begin{figure*}
  \centering
  \includegraphics[width=1\linewidth, angle=0,scale=1]{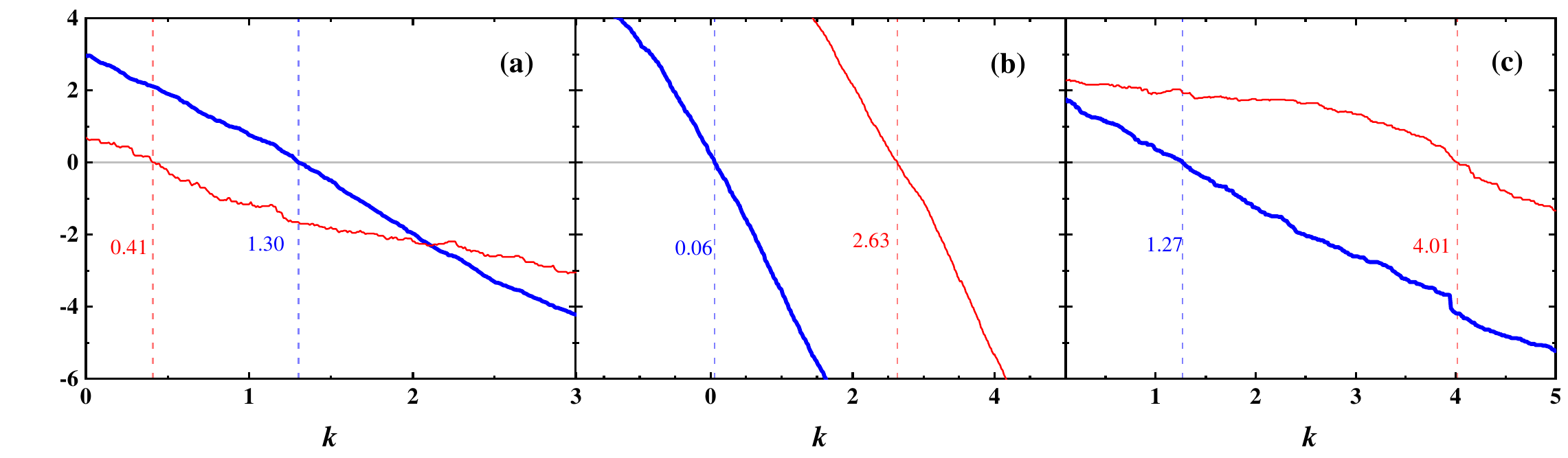}
  \caption{The $\tau$-statistic results of low-luminosity (thin solid line)
  and high-luminosity (thick solid line) GRBs for Samples Y15 (Panel a),
  BL (Panel b) and D22 (Panel c), respectively. The vertical lines show
  the expected $k$ values when $\tau$ is equal to zero.}
  \label{fig3}
\end{figure*}

Following our recent work, we utilize the Lynden Bell's c$^-$ method and
the $\tau$-statistic to determine the dependence of the peak luminosity on
the redshift, which is assumed to take the form
of $L\propto(1+z)^k$ \cite[see][for details]{2022MNRAS.513.1078D}.
By varying the $k$ value until $\tau$ reaches zero, we
obtain $k_l=0.41_{-0.4}^{+0.4}$ and $k_h=1.3_{-0.41}^{+0.36}$ in the Y15
sample, $k_l=2.63_{-0.3}^{+0.35}$ and $k_h=0.06_{-0.26}^{+0.28}$ in the
BL sample, $k_l=4.01_{-0.65}^{+0.65}$ and $k_h=1.27_{-0.63}^{+0.6}$ in the
D22 sample for low- and high-luminosity lGRBs, correspondingly. The results
are illustrated in Figure \ref{fig3}. It can be seen that the power-law
indices of $k_l$ and $k_h$ are significantly different in each sample.
Interestingly, we notice that $k_l$ is larger than $k_h$ in the two
samples of BL and D22, but the case is opposite for Sample Y15, which
may be caused by the lower proportion of low-luminosity lGRBs in the
Y15 sample. Although the derived $k$ values are sample-dependent,
they are roughly coincident with previous estimates. For
instance, \cite{2012MNRAS.423.2627W} derived the value of $k$
as $2.3_{-0.51}^{+0.56}$ for a sample of 95 lGRBs detected by
multiple satellites including Swift, Fermi/GBM, Konus-wind, and so
on. \cite{2017ApJ...850..161T} got a smaller value of $k\sim1.7$ for a
sample of 150 Konus-wind lGRBs with reliable redshift measurements. It
indicates that the $k$ values are less biased by the instrumental
effect as compared with the sample selection effect.

\section{GRB rate versus SFR}

\label{sec:results}

\begin{figure*}
  \centering
  \includegraphics[width=1.0\linewidth, angle=0,scale=1.0]{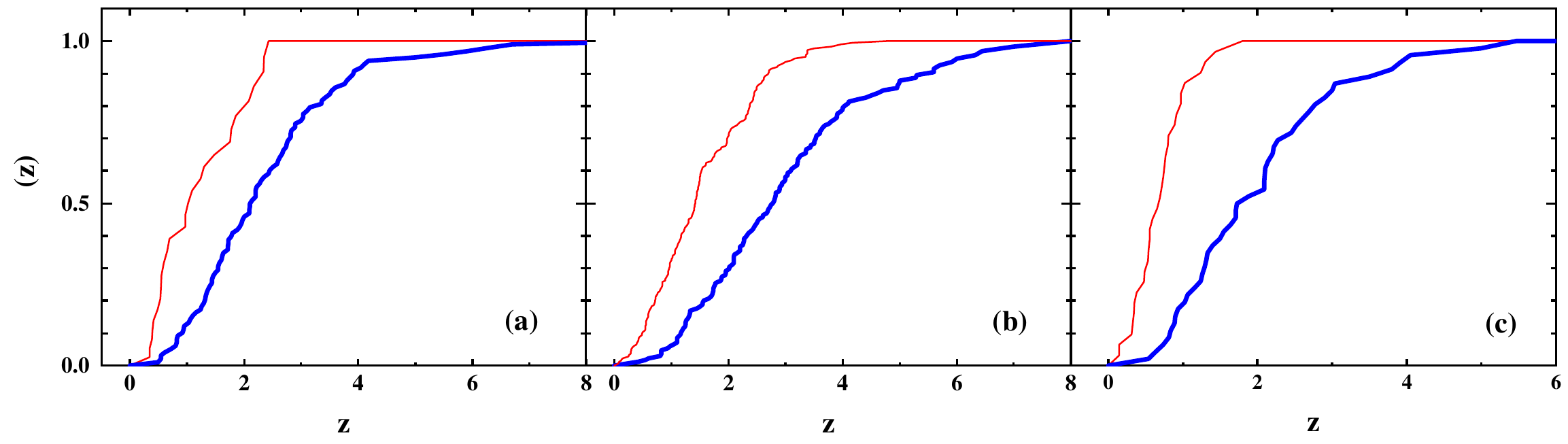}\\
  \caption{Cumulative redshift distributions of low (thin solid line)
  and high (thick solid line) luminosity lGRBs for the three samples.
  All data have been normalized to the maximum redshift. Panels (a),
  (b) and (c) correspond to the Y15, BL and D22 samples, respectively.}
  \label{fig4}
\end{figure*}

Using the cumulative redshift distribution function $\phi(z)$, one can
calculate the lGRB rate, $\rho(z)$, as done in D22. The normalized cumulative
redshift distributions of Samples Y15, BL and D22 are shown in Figure \ref{fig4},
where we find that the low-luminosity lGRBs are relatively closer to us than
those high-luminosity ones in each sample. In comparison, the cumulative
redshift distributions of low-luminosity lGRBs increase faster in the smaller
redshift domain. Interestingly, the median redshifts of both low- and high-luminosity
lGRBs in Sample Y15 are evidently larger than those corresponding medians of Sample
D22, but smaller than the medians of Sample BL. More interestingly, we find that
the median redshift of high-luminosity lGRBs in each sample is about two times larger
than that of low-luminosity lGRBs, in which the physical reason is unknown yet.
However, we can speculate that weak GRBs at high redshifts are most probably
low-luminosity bursts not detected by us due to the Malmquist bias
effect \citep{2021ApJ...914L..40D}. They need more effective ways to detect in
the future \citep{2021NatAs...5..262J,2021ExA....52..219T,2022PASA...39...32T}.

\begin{figure*}
\centering
\includegraphics[width=1.0\linewidth, angle=0,scale=1.0]{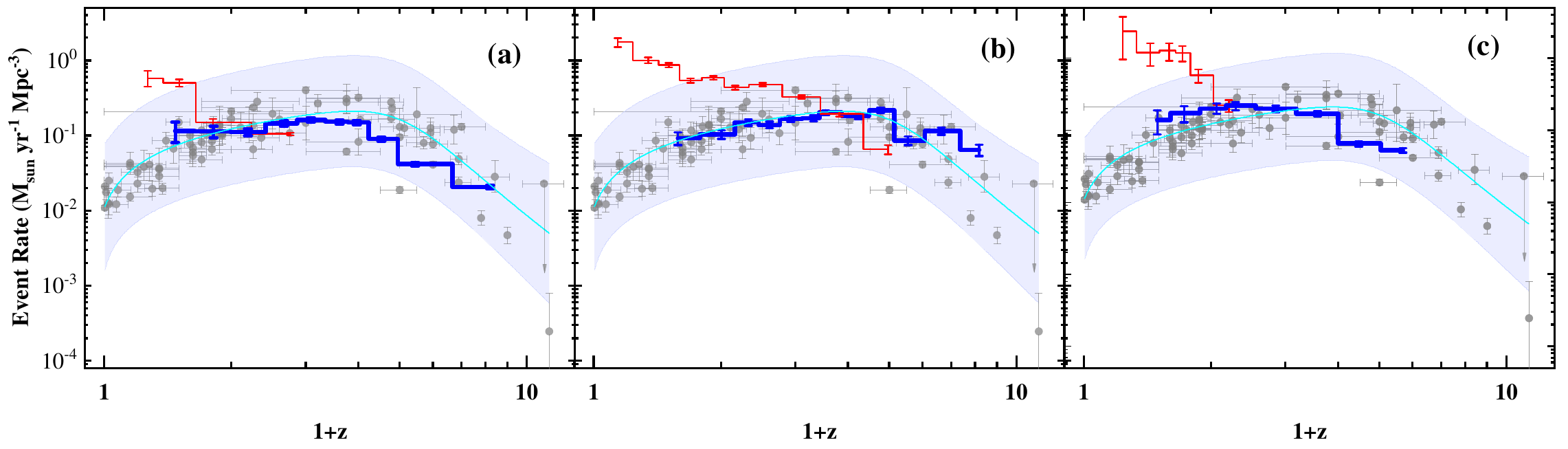} 
    \caption{Comparison between the SFR (gray points) and the event rates of
    low (red thin ladder lines) and high (blue thick ladder lines) luminosity
    lGRBs for Samples Y15 in Panel (a), BL in Panel (b) and D22 in Panel (c),
    respectively. The observed SFR data together with the best fitting line
    (cyan solid lines) have been taken from D22. The light shaded regions
    indicate the 3$\sigma$ range of SFR.}
  \label{fig5}
\end{figure*}

Now, we use the cumulative redshift distribution to compute the event
rates of the two classes of lGRBs in each sample. The results are compared
with the SFR in Figure \ref{fig5}. It can be obviously found that the
dependence of the event rate on redshift for the low- and high-luminosity
lGRBs is very different no matter whether the lGRB sample is complete or
not, which is consistent with our findings in D22. In other words, the
event rates of high-luminosity lGRBs match the SFR well in all redshift
ranges and the low-luminosity lGRBs exhibit significant excess at
low-redshift of $z<1$, particularly for the complete sample D22. Even
for the smaller and incomplete Y15 sample, the low-redshift excess is
still significant. On the other hand, Figure \ref{fig5} (b)
indicates that the event rate of a small fraction of low-luminosity lGRBs
with higher redshift of $z>1$ is somewhat in agreement with the SFR.
This is evidently different from the previous viewpoint that the
low-redshift excess was biased by the incompleteness of GRB
samples \citep[e.g.][]{2016A&A...587A..40P,2019MNRAS.488.4607L}.
Strictly speaking, the sample completeness can influence the estimation
of GRB rate slightly as displayed in Figure \ref{fig5}, but it would not
change the evolutionary trend of GRBs in nature and its importance may
be overestimated in some previous studies. Our results indicate that the
high-luminosity lGRBs are produced by the core-collapse of massive stars,
while most low-luminosity lGRBs could originate from other processes
including compact star mergers which are unrelated to the
star formation rate. Note that the limited GRB samples may bias
the estimation of GRB rate especially in the low-redshift region. To test
the reliability of our results, we have double-checked the data used in
Figure \ref{fig5}, ensuring that the number of data points in each
redshift bin is no less than 5. In this way, we still find that there is
an excesses at low redshift in both the complete and incomplete lGRB samples
based on Poisson statistics. The excess is mainly contributed by the low-luminosity lGRBs as well.

\section{Conclusions}

\label{sec:summary}


Based on the analysis of GRB rates for both complete and incomplete
lGRB samples, we draw the following conclusions:

(1) The relations between peak luminosity and redshift of lGRBs are
described by $L\propto(1+z)^k$, with  $k_l=0.41_{-0.4}^{+0.4}$
and $k_h=1.3_{-0.41}^{+0.36} $ for Sample Y15, $k_l=2.63_{-0.3}^{+0.35}$
and $k_h=0.06_{-0.26}^{+0.28}$ for Sample BL, $k_l=4.01_{-0.65}^{+0.65}$
and $k_h=1.27_{-0.63}^{+0.6} $ for Sample D22, respectively.

(2)	After dividing the lGRBs into low- and high-luminosity ones, we
find that the event rates of high-luminosity lGRBs match the SFR well
in all redshift ranges and the low-luminosity lGRBs show significant
excess at low-redshift of $z<1$. Therefore, we conclude that the
low-redshift excess of GRB rate is resulted from the low-luminosity lGRBs.

(3) We find that the lGRB rate always exceeds SFR at low redshift
of $z<1$ no matter whether the lGRB samples are complete or not.
That is to say, the low-redshift excess is intrinsically real and
it cannot be attributed to the sample incompleteness.

Consequently, we argue that the high- and low-luminosity lGRBs might
have different physical origins. High-luminosity bursts tracing the SFR
should be produced by the death of massive stars and the low-luminosity
bursts may come from other processes that is unrelated to the SFR.
The results might shed new light on the investigation of GRB progenitors
and central engines.

\section{Discussion}
\label{sec:discussion}

Currently, the origin of the low-luminosity GRBs is still an open question.
For example, \cite{2007ApJ...662.1111L} and \cite{2007ApJ...657L..73G} claimed
the existence of two physically distinct classes of lGRBs but cannot exclude
a single population. \cite{2021MNRAS.508...52L} showed that the low-luminosity
lGRBs are not the straightforward extension of high-luminosity ones through
the luminosity distribution of their sample. \cite{2022arXiv220814459P}
mentioned that a cocoon breakout is most likely the origin of low-luminosity
GRBs. Broad absorption lines produced in such a process have been observed
in several SNe associated with low-luminosity
lGRBs \cite[e.g.][]{2012ApJ...753...67B,2013ApJ...776...98X,2019Natur.565..324I}.
\cite{2007ApJ...657L..73G} investigated the GRB classification by considering
their probability of being associated with an SN Ib/c. They
found that low-luminosity GRBs are more likely to be associated with an SN Ib/c than
high-luminosity GRBs. However, it is still too difficult to infer the nature of
these two populations simply from their difference in the probability.
\cite{2012MNRAS.423.3049G} compared
the Type Ib/c event rate with the GRB rate. It is found that the local ratio of
GRB/SNe Ib/c is $<3\times10^{-3}$ in spirals, and is $<2\times10^{-2}$ in irregulars,
which means that only a tiny fraction of SNe Ib/c sources ends up by exploding as lGRBs.
Another possible origin involving the cocoon shock breakout from AGN disk was
also proposed \citep{2021ApJ...906L..11Z}. Particularly, \cite{2018A&A...616A.169M}
argued that the low-luminosity GRB 111005A represents an independent class of GRBs
differing from the typical core-collapse events and inferred that the GRB rate may
be biased towards low SFRs.

The discrepancy between the GRB rate and the SFR has been tried to be reconciled
by several groups by considering the cosmic metallicity
evolution \citep{2012ApJ...749...68S, 2019MNRAS.488.5823L} or an evolving luminosity
function of GRBs \citep{2002ApJ...574..554L, 2016A&A...587A..40P, 2021MNRAS.508...52L}.
\cite{2006Natur.442.1014S} derived the intrinsic rate of sub-energetic GRBs
as $\sim 230$ Gpc$^{-3}$\ yr$^{-1}$, which is about 10 times more abundant than
typical bright GRBs. The low-luminosity lGRBs are likely a unique population because
of the high local event rate and small beaming
factor \cite[see e.g.][]{2007ApJ...662.1111L,2007ApJ...657L..73G,2021ApJ...911L..19Z}.
However, the real reason of the event rate exceeding the SFR at low redshift remains
unknown. The most important finding in this study is that the low-redshift excess
of GRB rate is induced by low-luminosity events rather than faint lGRBs.
Moreover, the low-luminosity lGRBs at low redshift may not be related with
the death of massive stars, but should be most probably connected to the merger of compact stars
with a relatively low mass.

The reason that some previous works found no excess at low redshift
may be that their samples are mainly consisted of high-luminosity lGRBs \citep{2012MNRAS.423.2627W,2012ApJ...749...68S,2016A&A...587A..40P,2022arXiv220606390G},
since they have utilized the complete GRB samples such as the BAT6
sample \citep{2012ApJ...749...68S} which is dominated by bright Swift lGRBs.
\cite{2022MNRAS.513.1078D} also applied the same non-parametric method to
a complete sample including weak bursts and found the low-redshift excess
exists. Simultaneously, they verified that the GRB rate exceeds the SFR at
low redshift even for the non-complete GRB samples when the fraction of
low-redshift GRBs is sufficient enough. Interestingly, we find in this
study that the low-redshift excess is caused by the low-luminosity lGRBs
instead of the high-luminosity ones, which is independent of the
completeness of GRB samples.

An interesting issue is that the distributions of both luminosity and
redshift can be influenced by the detailed geometry of the jet. It is quite
clear that a realistic GRB jet could have some kinds of
structure \citep{2020ApJ...896...83G}. Therefore, how different jet models
affect the luminosity and redshift distributions is an important issue.
Some authors have tried to solve the problem by assuming a particular jet
profile in advance \citep[e.g.,][]{2004ApJ...606L..37N,2020ApJ...896...83G,2022A&A...661A.145B}.
For example, \cite{2004ApJ...606L..37N} discussed the influence of an universal
structured jet model on the redshift distribution. They also calculated the theoretical GRB rate
as a function of redshift and viewing angle. It is found that the structured jet model is
incompatible with the currently available observational data because of the redshift
selection effects. More recently, \cite{2022A&A...661A.145B} proposed that the
scatter of the luminosity in the co-moving frame can be influenced by the
angular emission profile of GRB jets. They also found that the mean luminosity
is insensitive to the on-axis luminosity of the jet. But still, they did not
find the actual effect of a particular structured jet model on the distributions
of luminosity and redshift. In practice, it is usually difficult to discriminate
whether the GRB ejecta is structured or isotropic through observations. It is
even more difficult to distinguish different types of structured jets.
In some cases, our line of sight may deviate from the jet
axis \citep{2023A&A...673A..20X}, which will lead to more difficulties in
deriving the luminosity function and event
rate of GRBs. We hope that the rapid increase of the number of GRBs
with redshift measured would bring more opportunity to us in the future.

\section*{Acknowledgements}

We thank the anonymous referee for helpful suggestions that led to an overall improvement of this study.
This work was supported by the National Natural Science Foundation of China (No. U2031118),
the Youth Innovations and Talents Project of Shandong Provincial Colleges and Universities
(Grant No. 201909118) and the Natural Science Foundations (ZR2018MA030, XKJJC201901).
YFH is supported by the National Key R\&D Program of China (2021YFA0718500),
by National SKA Program of China (No. 2020SKA0120300), by the National Natural Science
Foundation of China (Grant Nos. 12233002, 12041306, U1938201).

\bibliographystyle{aasjournal}
\bibliography{refs}

\begin{thebibliography}{}
\makeatletter
\relax
\def\mn@urlcharsother{\let\do\@makeother \do\$\do\&\do\#\do\^\do\_\do\%\do\~}
\def\mn@doi{\begingroup\mn@urlcharsother \@ifnextchar [ {\mn@doi@}
  {\mn@doi@[]}}
\def\mn@doi@[#1]#2{\def\@tempa{#1}\ifx\@tempa\@empty \href
  {http://dx.doi.org/#2} {doi:#2}\else \href {http://dx.doi.org/#2} {#1}\fi
  \endgroup}
\def\mn@eprint#1#2{\mn@eprint@#1:#2::\@nil}
\def\mn@eprint@arXiv#1{\href {http://arxiv.org/abs/#1} {{\tt arXiv:#1}}}
\def\mn@eprint@dblp#1{\href {http://dblp.uni-trier.de/rec/bibtex/#1.xml}
  {dblp:#1}}
\def\mn@eprint@#1:#2:#3:#4\@nil{\def\@tempa {#1}\def\@tempb {#2}\def\@tempc
  {#3}\ifx \@tempc \@empty \let \@tempc \@tempb \let \@tempb \@tempa \fi \ifx
  \@tempb \@empty \def\@tempb {arXiv}\fi \@ifundefined
  {mn@eprint@\@tempb}{\@tempb:\@tempc}{\expandafter \expandafter \csname
  mn@eprint@\@tempb\endcsname \expandafter{\@tempc}}}


\bibitem[\protect\citeauthoryear{Banerjee \& Guetta}{2022}]{2022A&A...661A.145B} Banerjee S., Guetta D., 2022, A\&A, 661, A145. doi:10.1051/0004-6361/202142628.
\bibitem[\protect\citeauthoryear{Bufano et al.}{2012}]{2012ApJ...753...67B} Bufano F., Pian E., Sollerman J., Benetti S., Pignata G., Valenti S., Covino S., et al., 2012, ApJ, 753, 67. doi:10.1088/0004-637X/753/1/67
\bibitem[Bromberg et al.(2013)]{2013ApJ...764..179B} Bromberg, O., Nakar, E., Piran, T., et al.\ 2013, \apj, 764, 179. doi:10.1088/0004-637X/764/2/179
\bibitem[\protect\citeauthoryear{Cobb et al.}{2006}]{2006ApJ...645L.113C} Cobb B.~E., Bailyn C.~D., van Dokkum P.~G., Natarajan P., 2006, ApJL, 645, L113. doi:10.1086/506271
\bibitem[\protect\citeauthoryear{Daigne, Rossi \& Mochkovitch}{2006}]{2006MNRAS.372.1034D}Daigne, F., Rossi, E. M., Mochkovitch, R., 2006, \mnras, 372, 1034. doi:10.1111/j.1365-2966.2006.10837.x
 \bibitem[\protect\citeauthoryear{Dainotti et al.}{2018}]{Dainotti2018}Dainotti, M. G. \& Amati, L. 2018, PASP, 130, 051001
\bibitem[\protect\citeauthoryear{Dainotti, Petrosian \& Bowden}{2021}]{2021ApJ...914L..40D} Dainotti M.~G., Petrosian V., Bowden L., 2021, ApJL, 914, L40. doi:10.3847/2041-8213/abf5e4
\bibitem[\protect\citeauthoryear{Deng et al.}{2022}]{Deng2022} Deng Q., Zhang Z.-B., Li X.-J., et al.,\ 2022, \apj, 940, 5. doi:10.3847/1538-4357/ac9590
\bibitem[\protect\citeauthoryear{Dong et al.}{2022}]{2022MNRAS.513.1078D} Dong X.~F., Li X.~J., Zhang Z.~B., Zhang X.~L., 2022, MNRAS, 513, 1078. doi:10.1093/mnras/stac949
\bibitem[\protect\citeauthoryear{Efron \& Petrosian}{1992}]{1992ApJ...399..345E} Efron B., Petrosian V., 1992, ApJ, 399, 345. doi:10.1086/171931
\bibitem[\protect\citeauthoryear{Eichler et al.}{1989}]{1989Natur.340..126E}Eichler, D., Livio, M., Piran, T., and Schramm, D. N. 1989, \nat, 340, 126. doi:10.1038/340126a0
\bibitem[\protect\citeauthoryear{El{\'\i}as \& Mart{\'\i}nez}{2018}]{2018RMxAA..54..309E} El{\'\i}as M., Mart{\'\i}nez O., 2018, RMxAA, 54, 309
\bibitem[\protect\citeauthoryear{Gehrels et al.}{2004}]{2004ApJ...611.1005G} Gehrels N., Chincarini G., Giommi P., Mason K.~O., Nousek J.~A., Wells A.~A., White N.~E., et al., 2004, ApJ, 611, 1005. doi:10.1086/422091
\bibitem[\protect\citeauthoryear{Ghirlanda \& Salvaterra}{2022}]{2022arXiv220606390G} Ghirlanda G., Salvaterra R., 2022, \apj, accepted, arXiv:2206.06390
\bibitem[\protect\citeauthoryear{Grieco et al.}{2012}]{2012MNRAS.423.3049G} Grieco V., Matteucci F., Meynet G., Longo F., Della Valle M., Salvaterra R., 2012, MNRAS, 423, 3049. doi:10.1111/j.1365-2966.2012.21052.x
\bibitem[Gruber et al.(2014)]{2014ApJS..211...12G} Gruber, D., Goldstein, A., Weller von Ahlefeld, V., et al.\ 2014, \apjs, 211, 12. doi:10.1088/0067-0049/211/1/12
\bibitem[\protect\citeauthoryear{Guetta \& Piran}{2005}]{2005A&A...435..421G} Guetta D., Piran T., 2005, A\&A, 435, 421. doi:10.1051/0004-6361:20041702
\bibitem[\protect\citeauthoryear{Guetta \& Della Valle}{2007}]{2007ApJ...657L..73G} Guetta D., Della Valle M., 2007, ApJL, 657, L73. doi:10.1086/511417
\bibitem[\protect\citeauthoryear{Guo et al.}{2020}]{2020ApJ...896...83G} Guo Q., Wei D.-M., Wang Y.-Z., Jin Z.-P., 2020, \apj, 896, 83. doi:10.3847/1538-4357/ab8f9d
\bibitem[\protect\citeauthoryear{Hjorth et al.}{2003}]{2003Natur.423..847H} Hjorth J., Sollerman J., M{\o}ller P., Fynbo J.~P.~U., Woosley S.~E., Kouveliotou C., Tanvir N.~R., et al., 2003, Natur, 423, 847. doi:10.1038/nature01750
\bibitem[\protect\citeauthoryear{Hopkins \& Beacom}{2006}]{2006ApJ...651..142H} Hopkins, A. M., Beacom, J. F., 2006, \apj, 651, 142. doi:
10.1086/506610
\bibitem[\protect\citeauthoryear{Izzo et al.}{2019}]{2019Natur.565..324I} Izzo L., de Ugarte Postigo A., Maeda K., Th{\"o}ne C.~C., Kann D.~A., Della Valle M., Sagues Carracedo A., et al., 2019, Natur, 565, 324. doi:10.1038/s41586-018-0826-3
\bibitem[\protect\citeauthoryear{Jakobsson et al.}{2006}]{2006A&A...447..897J}Jakobsson, P., Levan, A., Fynbo, J. P. U. et al., 2006, \aap, 447, 897. doi:10.1051/0004-6361:20054287
\bibitem[\protect\citeauthoryear{Jiang et al.}{2021}]{2021NatAs...5..262J} Jiang L., Wang S., Zhang B., Kashikawa N., Ho L.~C., Cai Z., Egami E., et al., 2021, NatAs, 5, 262. doi:10.1038/s41550-020-01266-z
\bibitem[\protect\citeauthoryear{Kistler et al.}{2008}]{2008ApJ...673L.119K}Kistler, M. D., Y\"{u}ksel, H., Beacom, J. F., Stanek, K. Z., 2008, \apj, 673, 119. doi:10.1086/527671
\bibitem[\protect\citeauthoryear{Kouveliotou et al.}{1993}]{1993ApJ...413L.101K} Kouveliotou C., Meegan C.~A., Fishman G.~J., Bhat N.~P., Briggs M.~S., Koshut T.~M., Paciesas W.~S., et al., 1993, ApJL, 413, L101. doi:10.1086/186969
\bibitem[\protect\citeauthoryear{Lan et al.}{2019}]{2019MNRAS.488.4607L} Lan G.-X., Zeng H.-D., Wei J.-J., Wu X.-F., 2019, MNRAS, 488, 4607. doi:10.1093/mnras/stz2011
\bibitem[\protect\citeauthoryear{Lan et al.}{2021}]{2021MNRAS.508...52L} Lan G.-X., Wei J.-J., Zeng H.-D., Li Y., Wu X.-F., 2021, MNRAS, 508, 52. doi:10.1093/mnras/stab2508
\bibitem[\protect\citeauthoryear{Le \& Dermer}{2007}]{2007ApJ...661..394L}Le T., Dermer, C. D., 2007, \apj, 661, 394. doi:10.1086/513460
\bibitem[\protect\citeauthoryear{Le, Ratke \& Mehta}{2020}]{2020MNRAS.493.1479L} Le T., Ratke C., Mehta V., 2020, MNRAS, 493, 1479. doi:10.1093/mnras/staa366
\bibitem[\protect\citeauthoryear{Li et al.}{2023}]{2023MNRAS.tmp.1633L} Li Q.~M., Zhang Z.~B., Han X.~L., Zhang K.~J., Xia X.~L., Hao C.~T., 2023, MNRAS, 524, 1096. doi:10.1093/mnras/stad1648
\bibitem[\protect\citeauthoryear{Liang et al.}{2007}]{2007ApJ...662.1111L} Liang E., Zhang B., Virgili F., Dai Z.~G., 2007, ApJ, 662, 1111. doi:10.1086/517959
\bibitem[\protect\citeauthoryear{Lloyd-Ronning, Aykutalp \& Johnson}{2019}]{2019MNRAS.488.5823L} Lloyd-Ronning N.~M., Aykutalp A., Johnson J.~L., 2019, MNRAS, 488, 5823. doi:10.1093/mnras/stz2155
\bibitem[\protect\citeauthoryear{Lloyd-Ronning, Fryer, \& Ramirez-Ruiz}{2002}]{2002ApJ...574..554L} Lloyd-Ronning N.~M., Fryer C.~L., Ramirez-Ruiz E., 2002, ApJ, 574, 554. doi:10.1086/341059
\bibitem[\protect\citeauthoryear{Lynden-Bell}{1971}]{1971MNRAS.155...95L} Lynden-Bell D., 1971, MNRAS, 155, 95. doi:10.1093/mnras/155.1.95
\bibitem[\protect\citeauthoryear{Madau et al.}{1998}]{1998ApJ...498..106M}Madau, P., Pozzetti, L., and Dickinson, M. 1998, \apj, 498, 106. doi:10.1086/305523
\bibitem[\protect\citeauthoryear{Micha{\l}owskI et al.}{2018}]{2018A&A...616A.169M} Micha{\l}owskI M.~J., Xu D., Stevens J., et al., 2018, A\&A, 616, A169. doi:10.1051/0004-6361/201629942
\bibitem[\protect\citeauthoryear{Narayan, Paczy{\'n}ski \& Piran}{1992}]{1992ApJ...395L..83N}Narayan, R., Paczy{\'n}ski, B., and Piran, T., 1992, \apjl, 395, L83. doi: 10.1086/186493
\bibitem[\protect\citeauthoryear{Nakar, Granot, \& Guetta}{2004}]{2004ApJ...606L..37N}Nakar E., Granot J., Guetta D., 2004, ApJL, 606, L37. doi:10.1086/421107
\bibitem[\protect\citeauthoryear{Nugent et al.}{2022}]{2022arXiv220601764N} Nugent A.~E., Fong W.-F., Dong Y., Leja J., Berger E., Zevin M., Chornock R., et al., 2022, arXiv, arXiv:2206.01764
\bibitem[\protect\citeauthoryear{Paczy{\'n}ski}{1991}]{Paczynski1991}Paczy{\'n}ski B., 1991, Acta Astron., 41, 257.
\bibitem[\protect\citeauthoryear{Paczy{\'n}ski}{1998}]{1998ApJ...494L..45P}Paczy{\'n}ski B., 1998, ApJL, 494, L45. doi:10.1086/311148
\bibitem[\protect\citeauthoryear{Pais, Piran \& Nakar}{2023}]{2022arXiv220814459P} Pais M., Piran T., Nakar E., 2023, \mnras, 519, 1941
\bibitem[\protect\citeauthoryear{Palmerio \& Daigne}{2021}]{2021A&A...649A.166P} Palmerio J.~T., Daigne F., 2021, A\&A, 649, A166. doi:10.1051/0004-6361/202039929
\bibitem[\protect\citeauthoryear{Pescalli et al.}{2016}]{2016A&A...587A..40P} Pescalli A., Ghirlanda G., Salvaterra R., Ghisellini G., Vergani S.~D., Nappo F., Salafia O.~S., et al., 2016, A\&A, 587, A40. doi:10.1051/0004-6361/201526760
\bibitem[\protect\citeauthoryear{Petrosian, Kitanidis \& Kocevski}{2015}]{2015ApJ...806...44P} Petrosian V., Kitanidis E., Kocevski D., 2015, ApJ, 806, 44. doi:10.1088/0004-637X/806/1/44
\bibitem[\protect\citeauthoryear{Petrosian \& Dainotti}{2023}]{2023arXiv230515081V} Petrosian V., Dainotti M.~G., 2023, arXiv, arXiv:2305.15081. doi:10.48550/arXiv.2305.15081
\bibitem[\protect\citeauthoryear{Salvaterra et al.}{2012}]{2012ApJ...749...68S} Salvaterra R., Campana S., Vergani S.~D., Covino S., D'Avanzo P., Fugazza D., Ghirlanda G., et al., 2012, ApJ, 749, 68. doi:10.1088/0004-637X/749/1/68
\bibitem[\protect\citeauthoryear{Sazonov, Lutovinov \& Sunyaev}{2004}]{2004Natur.430..646S} Sazonov, S. Yu., Lutovinov, A. A., Sunyaev, R. A., \nat, 2004, 430, 646. doi:https://doi.org/10.1038/nature02748
\bibitem[\protect\citeauthoryear{Soderberg et al.}{2006}]{2006Natur.442.1014S} Soderberg A.~M., Kulkarni S.~R., Nakar E., Berger E., Cameron P.~B., Fox D.~B., Frail D., et al., 2006, Natur, 442, 1014. doi:10.1038/nature05087
\bibitem[\protect\citeauthoryear{Stanek et al.}{2003}]{2003ApJ...591L..17S} Stanek K.~Z., Matheson T., Garnavich P.~M., Martini P., Berlind P., Caldwell N., Challis P., et al., 2003, ApJL, 591, L17. doi:10.1086/376976
\bibitem[\protect\citeauthoryear{Tang et al.}{2019}]{2019ApJS..245....1T}Tang, C.-H., Huang, Y.-F., Geng, J.-J. and Zhang, Z.-B., 2019, \apjs, 245, 1.
\bibitem[\protect\citeauthoryear{Tanvir et al.}{2021}]{2021ExA....52..219T} Tanvir N.~R., Le Floc'h E., Christensen L., Caruana J., Salvaterra R., Ghirlanda G., Ciardi B., et al., 2021, ExA, 52, 219. doi:10.1007/s10686-021-09778-w
\bibitem[Tarnopolski(2019a)]{2019ApJ...870..105T} Tarnopolski, M.\ 2019, \apj, 870, 105. doi:10.3847/1538-4357/aaf1c5
\bibitem[Tarnopolski(2019b)]{2019ApJ...887...97T} Tarnopolski, M.\ 2019, \apj, 887, 97. doi:10.3847/1538-4357/ab4fe6
\bibitem[Tarnopolski(2022)]{2022A&A...657A..13T} Tarnopolski, M.\ 2022, \aap, 657, 13. doi:10.1051/0004-6361/202038645
\bibitem[\protect\citeauthoryear{Thomas et al.}{2022}]{2022PASA...39...32T} Thomas M., Trenti M., Greiner J., Skrutskie M., Forbes D.~A., Klose S., Mack K.~J., et al., 2022, PASA, 39, e032. doi:10.1017/pasa.2022.22
\bibitem[\protect\citeauthoryear{Tsvetkova et al.}{2017}]{2017ApJ...850..161T} Tsvetkova A., Frederiks D., Golenetskii S., Lysenko A., Oleynik P., Pal'shin V., Svinkin D., et al., 2017, ApJ, 850, 161. doi:10.3847/1538-4357/aa96af
\bibitem[\protect\citeauthoryear{Virgili et al.}{2011}]{2011MNRAS.417.3025V}Virgili, F. J., Zhang, B., Nagamine, K., Choi, J. H., 2011, \mnras, 417, 3026. doi:10.1111/j.1365-2966.2011.19459.x
\bibitem[\protect\citeauthoryear{Wanderman \& Piran}{2010}]{2010MNRAS.406.1944W} Wanderman, D., Piran, T., 2010, \mnras, 406, 1944. doi:10.1111/j.1365-2966.2010.16787.x
\bibitem[\protect\citeauthoryear{Wanderman \& Piran}{2015}]{2015MNRAS.448.3026W} Wanderman D., Piran T., 2015, MNRAS, 448, 3026. doi:10.1093/mnras/stv123
\bibitem[\protect\citeauthoryear{Wang \& Dai}{2009}]{2009MNRAS.400L..10W} Wang F.~Y., Dai Z.~G., 2009, MNRAS, 400, L10. doi:10.1111/j.1745-3933.2009.00751.x
\bibitem[\protect\citeauthoryear{Wang et al.}{2020}]{2020ApJ...893...77W} Wang F.-F., Zou Y.-C., Liu F. X. et al., 2020, \apj, 893, 77. doi:10.3847/1538-4357/ab0a86
\bibitem[\protect\citeauthoryear{Woosley}{1993}]{1993AAS...182.5505W} Woosley S.~E., 1993, \aas, 182, 5505
\bibitem[\protect\citeauthoryear{Woosley \& Bloom}{2006}]{2006ARA&A..44..507W} Woosley S.~E., Bloom J.~S., 2006, ARA\&A, 44, 507. doi:10.1146/annurev.astro.43.072103.150558
\bibitem[\protect\citeauthoryear{Wu et al.}{2012}]{2012MNRAS.423.2627W} Wu S.-W., Xu D., Zhang F.-W., Wei D.-M., 2012, MNRAS, 423, 2627. doi:10.1111/j.1365-2966.2012.21068.x
\bibitem[\protect\citeauthoryear{Xu et al.}{2013}]{2013ApJ...776...98X} Xu D., de Ugarte Postigo A., Leloudas G., Kr{\"u}hler T., Cano Z., Hjorth J., Malesani D., et al., 2013, ApJ, 776, 98. doi:10.1088/0004-637X/776/2/98
\bibitem[\protect\citeauthoryear{Xu et al.}{2023}]{2023A&A...673A..20X} Xu F., Huang Y.-F., Geng J.-J., et al.,  2023, A\&A, 673, A20. doi:10.1051/0004-6361/202245414
\bibitem[\protect\citeauthoryear{Yu et al.}{2015}]{2015ApJS..218...13Y} Yu H., Wang F.~Y., Dai Z.~G., Cheng K.~S., 2015, ApJS, 218, 13. doi:10.1088/0067-0049/218/1/13
\bibitem[\protect\citeauthoryear{Y{\"u}ksel et al.}{2008}]{2008ApJ...683L...5Y} Y{\"u}ksel H., Kistler M.~D., Beacom J.~F., Hopkins A.~M., 2008, ApJL, 683, L5. doi:10.1086/591449
\bibitem[\protect\citeauthoryear{Zhang et al.}{2009}]{2009ApJ...703.1696Z} Zhang B., Zhang B. B., Virgili F. J., et al., 2009, \apj, 703, 1696. doi: 10.1088/0004-637X/703/2/1696
\bibitem[\protect\citeauthoryear{Zhang}{2018}]{2018pgrb.book.....Z} Zhang B., 2018, The Physics of Gamma-Ray Bursts. Cambridge Univ. Press, New York, doi:10.1017/9781139226530
\bibitem[\protect\citeauthoryear{Zhang \& Wang}{2018}]{2018ApJ...852....1Z} Zhang G.~Q., Wang F.~Y., 2018, ApJ, 852, 1. doi:10.3847/1538-4357/aa9ce5
\bibitem[\protect\citeauthoryear{Zhang et al.}{2021}]{2021MNRAS.503.3262Z}Zhang, K., Zhang, Z. B., Huang, Y. F., et al., 2021, \mnras, 503, 3262.
\bibitem[\protect\citeauthoryear{Zhang et al.}{2020a}]{2020RAA....20..201Z} Zhang, X. L., Zhang, C. T., Li, X. J., et al., 2020a, RAA, 20, 201. doi:10.1088/1674-4527/20/12/201
\bibitem[\protect\citeauthoryear{Zhang \& Choi}{2008}]{2008A&A...484..293Z} Zhang Z.-B., Choi C.-S., 2008, A\&A, 484, 293. doi:10.1051/0004-6361:20079210
\bibitem[Zhang et al.(2016)]{2016MNRAS.462.3243Z} Zhang, Z. B., Yang, E.-B., Choi, C.-S., et al.\ 2016, \mnras, 462, 3243. doi:10.1093/mnras/stw1835
\bibitem[\protect\citeauthoryear{Zhang et al.}{2018}]{2018PASP..130e4202Z} Zhang Z.~B., Zhang C.~T., Zhao Y.~X., Luo J.~J., Jiang L.~Y., Wang X.~L., Han X.~L., et al., 2018, PASP, 130, 054202. doi:10.1088/1538-3873/aaa6af
\bibitem[Zhang et al. (2020b)]{2020ApJ...902...40Z} Zhang, Z. B., Jiang, M., Zhang, Y., et al.\ 2020b, \apj, 902, 40. doi:10.3847/1538-4357/abb400
\bibitem[\protect\citeauthoryear{Zhu et al.}{2021}]{2021ApJ...911L..19Z} Zhu J.-P., Wang K., Zhang B., Yang Y.-P., Yu Y.-W., Gao H., 2021, ApJL, 911, L19. doi:10.3847/2041-8213/abf2c3
\bibitem[\protect\citeauthoryear{Zhu et al.}{2021}]{2021ApJ...906L..11Z} Zhu J.-P., Zhang B., Yu Y.-W., Gao H., 2021, ApJL, 906, L11. doi:10.3847/2041-8213/abd412









\makeatother
\end{thebibliography}

\end{document}